\documentclass[fleqn, usenatbib]{mnras}
\usepackage{newtxtext,newtxmath}
\usepackage[T1]{fontenc}
\usepackage[utf8]{inputenc}
\usepackage{graphicx}
\usepackage{stfloats}
\usepackage{xcolor}
\usepackage{amsmath}
\usepackage{commath}
\usepackage{siunitx}
\usepackage{nth}
\usepackage[capitalise, noabbrev]{cleveref}
\usepackage{enumitem}
\usepackage{csvsimple}
\hypersetup{unicode}

\crefname{equation}{equation}{equations}

\DeclareRobustCommand{\VAN}[3]{#2}
\let\VANthebibliography\thebibliography
\def\thebibliography{\DeclareRobustCommand{\VAN}[3]{##3}\VANthebibliography}

\bibpunct[]{(}{)}{;}{a}{}{,}

\newcommand{\program}{\textsc}
\newcommand{\ssim}{\sim \!}

\title[Dust properties at the earliest epochs]{An empirical study of dust properties at the earliest epochs}

\author[J. Witstok et al.]{{Joris Witstok$^{1,2}$\thanks{E-mail: \href{mailto:jnw30@cam.ac.uk}{jnw30@cam.ac.uk}},
    Gareth C. Jones$^{3}$, Roberto Maiolino$^{1,2,4}$\thanks{E-mail: \href{mailto:rm665@cam.ac.uk}{rm665@cam.ac.uk}}, Renske Smit$^{5}$, and 
    }
    \newauthor{Raffaella Schneider$^{6,7,8}$
    }
    \\
    $^{1}$Kavli Institute for Cosmology, University of Cambridge, Madingley Road, Cambridge CB3 0HA, UK
    \\
    $^{2}$Cavendish Laboratory, University of Cambridge, 19 JJ Thomson Avenue, Cambridge CB3 0HE, UK
    \\
    $^{3}$Department of Physics, University of Oxford, Denys Wilkinson Building, Keble Road, Oxford OX1 3RH, UK
    \\
    $^{4}$Department of Physics and Astronomy, University College London, Gower Street, London WC1E 6BT, UK
    \\
    $^{5}$Astrophysics Research Institute, Liverpool John Moores University, 146 Brownlow Hill, Liverpool L3 5RF, UK
    \\
    $^{6}$Dipartimento di Fisica, Sapienza, Universit\`{a} di Roma, Piazzale Aldo Moro 5, 00185 Roma, Italy
    \\
    $^{7}$INFN, Sezione di Roma I, Piazzale Aldo Moro 2, 00185 Roma, Italy
    \\
    $^{8}$INAF/Osservatorio Astronomico di Roma, Via di Frascati 33, 00040 Monte Porzio Catone, Italy
}

\date{Accepted ---. Received ---; in original form ---}

\pubyear{2022}

\AtBeginDocument{
  \hypersetup{
    pdftitle={An empirical study of dust properties at the earliest epochs},
    pdfauthor={J. Witstok et al.},
    pdfsubject={An empirical analysis of the properties of the dust continuum emission in a sample of 17 galaxies in the early Universe ($4 < z < 8$) with well-sampled FIR SEDs compiled from the literature},
    pdfkeywords={{galaxies: high-redshift} -- {dark ages, reionization, first stars} -- {methods: observational} -- {ISM: dust, extinction}}
  }
}

\begin{document}
\label{firstpage}
\pagerange{\pageref{firstpage}--\pageref{lastpage}}
\maketitle

\begin{abstract}
    We present an empirical analysis of the properties of dust-continuum emission in a sample of 17 galaxies in the early Universe ($4 < z < 8$) with well-sampled far-infrared (FIR) spectral energy distributions (SEDs) compiled from the literature. We place our results into context by self-consistently comparing to samples of nearby star-forming galaxies, luminous infrared galaxies (LIRGs), and quasars. With the exception of two sources, we find no significant evolution in the dust emissivity index across cosmic time, measuring a consistent value of $\beta_\text{IR} = 1.8 \pm 0.3$ at $z > 4$, suggesting the effective dust properties do not change dramatically for most galaxies. Despite having comparable stellar masses, we find the high-redshift galaxies to be similar to, or even more extreme than, LIRGs in the HERUS sample in terms of dust temperature ($T_\text{dust} > 40 \, \mathrm{K}$) and IR luminosity ($L_\text{IR} > 10^{11} \, \mathrm{L_\odot}$). We find the dust temperature evolves mildly towards high redshift, though the LIRGs and quasars exhibit elevated temperatures indicating a more efficient and/or additional heating mechanism. Where available, we compare stellar-mass estimates to our inferred dust masses, whose degeneracy with dust temperature can only be mitigated with a well-constrained SED. In merely half of the cases the dust yield may be explained by supernovae alone, with four sources ($44\%$) significantly exceeding a highly optimistic yield where $M_\text{dust} \approx 0.01 M_*$. We discuss possible explanations for this apparent inconsistency and potential observational biases in the measurements of the dust properties of high-redshift galaxies, including in the current IR-bright sample.
\end{abstract}


\section{Introduction}
\label{sec:Introduction}

Cosmic dust grains are a prominent agent in the physical processes governing galaxy formation and evolution on the scale of the interstellar medium (ISM). Dust catalyses the formation of molecules \citep{2017MolAs...9....1W, 2018MNRAS.474.1545C} and the fragmentation of gas clouds \citep{2005ApJ...626..627O, 2006MNRAS.369.1437S}, two mechanisms that are essential to star formation. Furthermore, while the dust mass of a galaxy is negligible compared to its stellar or gas mass \citep{1998ApJ...501..643D, 2007ApJ...663..866D, 2011A&A...533A..16W}, dust grains absorb a significant fraction of the optical and ultraviolet (UV) light and in the infrared (IR) thermally re-emit the absorbed energy \citep[e.g.][]{1979ARA&A..17...73S, 1989ESASP.290...93D, 1999ApJ...521...64M, 2000ApJ...533..682C, 2001ApJ...548..296W, 2003ARA&A..41..241D}. This process has been shown to occur even among the first galaxies to emerge in the Epoch of Reionisation \citep[EoR; e.g.][]{2017ApJ...851...40L, 2019ApJ...874...27T, 2020MNRAS.493.4294B, 2021Natur.597..489F}.

A general consensus has been established on the various channels via which dust is formed across cosmic time. Specifically, the main sites of dust creation are thought to be asymptotic giant branch (AGB) stars, supernova (SN) events, and grain growth in the ISM \citep[e.g.][]{2015MNRAS.451L..70M, 2020MNRAS.494.1071G, 2022MNRAS.512..989D}. However, it is still unclear what is the exact composition and abundance of dust, particularly in the early Universe, when the age of the Universe was comparable to typical dust formation timescales \citep{2001MNRAS.325..726T, 2007MNRAS.378..973B, 2019A&A...624L..13L, 2020MNRAS.497..956S, 2023arXiv230205468W}. In addition, we do not yet fully understand how grain growth occurs in the ISM. It has been shown that this requires very small grains, with sizes smaller than $10 \, \mathrm{nm}$, in the cold neutral medium \citep{2009ASPC..414..453D, 2016ApJ...831..147Z}, and that once the grains are incorporated in dense molecular clouds, their growth becomes problematic due to the formation of icy mantles \citep{2016MNRAS.463L.112F, 2018MNRAS.476.1371C}. As a result, there exists significant observational uncertainty on fundamental galaxy properties such as the star formation rate (SFR), especially when they are inferred exclusively from rest-frame UV and optical measurements \citep[e.g.][]{2014ARA&A..52..415M}. Indeed, recent works suggest a significant fraction of obscured star formation activity may be missed \citep{2020A&A...643A...4F, 2021Natur.597..489F, 2020A&A...643A...3S, 2022MNRAS.512...58F, 2022MNRAS.517.5930S}.

With the advent of the Atacama Large Millimeter/submillimeter Array (ALMA), the first statistical samples of galaxies in the EoR detected by their dust-continuum emission are starting to be assembled \citep[e.g.][]{2022ApJ...928...31S, 2022MNRAS.515.3126I}. Nearly all of these sources, however, are only observed in a single photometric band, making it difficult to retrieve properties of the dust \citep{2021MNRAS.503.4878S, 2022MNRAS.513.3122S}. Typically, the spectral energy distribution (SED) is modelled as a modified blackbody \citep[``greybody''; see][]{2022MNRAS.515.1751W}, assigning ensemble properties to the dust through the temperature and emissivity parameters, $T_\text{dust}$ and $\beta_\text{IR}$, which are commonly assumed for galaxies with few photometric detections in the far infrared (FIR).

In this work, instead, we compile a list of high-redshift sources with well-sampled FIR SEDs. We particularly focus on the dust emissivity power-law index $\beta_\text{IR}$, which is connected to the microscopic properties of grains and therefore potentially holds the key to uncovering their evolutionary pathways \citep[e.g.][]{2013A&A...558A..62J}. Furthermore, we investigate the dust mass and temperature, which become degenerate for poorly sampled SEDs, but have the potential of revealing the main origin of a galaxy's dust content built up over relatively short timescales \citep[e.g.][]{2015A&A...577A..80M, 2020MNRAS.494.1071G, 2022MNRAS.512..989D}. In \cref{sec:Methods}, we briefly describe this sample and two other samples of galaxies in the local Universe. \Cref{sec:Results_and_discussion} describes the results of our analysis and discusses them in light of dust evolutionary mechanisms and potential changes in the properties of dust across cosmic time. \Cref{sec:Summary} provides the conclusions of our findings. We adopt the cosmological parameters $\Omega_\text{m} = 0.3$, $\Omega_\Lambda = 0.7$, and $H_0 = 70 \, \mathrm{km \, s^{-1} \, Mpc^{-1}}$ throughout.

\begingroup
    \setlength{\tabcolsep}{10pt} 
    \renewcommand{\arraystretch}{1.2} 
    \begin{table*}
        \centering
        \caption
        {Properties of local-Universe galaxy samples considered in this work.}
        \begin{tabular}{llllllp{5cm}}
            Sample & Galaxy type & $z$ & $M_* \, (\mathrm{M_\odot})$ & $A_\text{dust} \, (\mathrm{kpc^2})$ & $n_\text{phot}$ & References \\
            \hline
            JINGLE & Star-forming & $0.028_{-0.006}^{+0.011}$ & $1.3_{-1.0}^{+2.9} \cdot 10^{10}$ & $1.3_{-0.8}^{+1.9} \cdot 10^{2}$ & $5$ & \citet{2018MNRAS.481.3497S}, \citet{2019MNRAS.486.4166S}, \citet{2019MNRAS.489.4389L} \\
            HERUS & (U)LIRG & $0.086_{-0.043}^{+0.048}$ & $9.0_{-5.0}^{+9.0} \cdot 10^{10}$ & $2.2_{-1.9}^{+2.5} \cdot 10^{3}$ & $5$ & \citet{2003AJ....126.1607S}, \citet{2018MNRAS.475.2097C} \\
            PG & QSO & $0.14_{-0.08}^{+0.19}$ & -- & -- & $6$ & \citet{2015ApJS..219...22P} \\
        \end{tabular}
        \flushleft
        \textbf{Notes.} Listed properties are their type (see \cref{ssec:Methods:Samples_in_the_local_Universe}), redshift ($z$), stellar mass ($M_*$) and area of the dust emission ($A_\text{dust}$), number of photometric detections used for the fitting routine ($n_\text{phot}$), and references to the relevant works describing the survey. Quantities quoted are median values with error bars reflecting the \nth{16} and \nth{84} percentiles.
        \label{tab:Local_sources}
    \end{table*}
\endgroup
\begingroup
    \setlength{\tabcolsep}{10pt} 
    \renewcommand{\arraystretch}{1.2} 
    \begin{table*}
        \centering
        \caption{Properties of high-redshift sources considered in this work.}
        \begin{tabular}{lllllll}
            Source & Type & $z$ & $M_* \, (\mathrm{M_\odot})$ & $A_\text{dust} \, (\mathrm{kpc^2})$ & $n_\text{phot}$ & References \\
            \hline
            \begin{filecontents*}{FIR_SED_results.csv}
object|type|z|Mstar|Ad|nphot|citations|Md|Td|Tp|betaIR|lam|LIR
GN20|SMG|4.0553|$1.1_{-0.4}^{+0.6} \cdot 10^{11}$|$50.3$|6|(\hyperlink{cite.2006MNRAS.370.1185P}{4}), (\hyperlink{cite.2011ApJ...740L..15M}{8}), (\hyperlink{cite.2014A&A...569A..98T}{12})|$358_{-47}^{+53}$|$37_{-1}^{+1}$|$34_{-1}^{+1}$|$1.9_{-0.1}^{+0.1}$|$55_{ -4 }^{ +4 }$|$152_{-13}^{+13}$
GN20.2b|SMG|4.0563|$1.1_{-0.4}^{+0.6} \cdot 10^{11}$|$31.4$|4|(\hyperlink{cite.2014A&A...569A..98T}{12})|$127_{-40}^{+117}$|$39_{-10}^{+5}$|$36_{-9}^{+4}$|$1.8_{-0.2}^{+0.2}$|$39_{ -7 }^{ +17 }$|$69_{-45}^{+41}$
ID141|SMG|4.243|nan|$4.8$|10|(\hyperlink{cite.2011ApJ...740...63C}{7}), (\hyperlink{cite.2020ApJ...898...33C}{26})|$173_{-11}^{+12}$|$52_{-1}^{+1}$|$35_{-1}^{+1}$|$1.91_{-0.07}^{+0.07}$|$128_{ -4 }^{ +5 }$|$163_{-8}^{+8}$
MORA-3|SFG|4.63|nan|$16.6$|9|(\hyperlink{cite.2007ApJ...671.1531Y}{5}), (\hyperlink{cite.2019ApJ...877...45M}{22}), (\hyperlink{cite.2020ApJ...890..171J}{28}), (\hyperlink{cite.2013MNRAS.436.1919C}{9}), (\hyperlink{cite.2021ApJ...923..215C}{34})|$138_{-20}^{+25}$|$41_{-2}^{+2}$|$36_{-2}^{+2}$|$1.8_{-0.1}^{+0.1}$|$55_{ -5 }^{ +5 }$|$89_{-12}^{+13}$
GN10|SFG|5.303|$1.2_{-0.1}^{+0.1} \cdot 10^{11}$|$0.8$|11|(\hyperlink{cite.2020ApJ...895...81R}{31})|$81_{-11}^{+13}$|$80_{-2}^{+2}$|$45_{-1}^{+1}$|$2.7_{-0.2}^{+0.2}$|$201_{ -10 }^{ +11 }$|$176_{-20}^{+20}$
SPT0346-52|SFG|5.656|nan|$1.8$|17|(\hyperlink{cite.2016MNRAS.457.4406A}{15}), (\hyperlink{cite.2020MNRAS.498.4109J}{29})|$146_{-3}^{+4}$|$80.3_{-0.4}^{+0.4}$|$46_{-1}^{+0}$|$1.65_{-0.02}^{+0.02}$|$202_{ -3 }^{ +3 }$|$411_{-8}^{+8}$
MORA-4|SFG|5.85|$3.2_{-1.5}^{+1.0} \cdot 10^{9}$|$1.8$|10|(\hyperlink{cite.2019ApJ...887...55C}{21}), (\hyperlink{cite.2021ApJ...923..215C}{34})|$66_{-10}^{+11}$|$50_{-2}^{+2}$|$34_{-2}^{+3}$|$1.88_{-0.10}^{+0.10}$|$129_{ -11 }^{ +11 }$|$51_{-6}^{+9}$
J2310+1855|QSO|6.0031|$7.2 \cdot 10^{9}$|$4.5$|11|(\hyperlink{cite.2018A&A...619A..39F}{18}), (\hyperlink{cite.2019ApJ...876...99S}{24}), (\hyperlink{cite.2019MNRAS.489.3939C}{20}), (\hyperlink{cite.2020ApJ...889..162L}{30}), (\hyperlink{cite.2022A&A...665A.107T}{38})|$136_{-3}^{+3}$|$55.9_{-0.3}^{+0.3}$|$38.2_{-0.4}^{+0.3}$|$1.83_{-0.03}^{+0.03}$|$116_{ -1 }^{ +2 }$|$196_{-4}^{+4}$
J1319+0959|QSO|6.133|nan|$8.8$|7|(\hyperlink{cite.2017ApJ...845..138S}{17}), (\hyperlink{cite.2019MNRAS.489.3939C}{20})|$132_{-28}^{+30}$|$43_{-3}^{+4}$|$35_{-2}^{+3}$|$1.7_{-0.2}^{+0.2}$|$76_{ -10 }^{ +9 }$|$88_{-14}^{+19}$
J0100+2802|QSO|6.327|nan|$11.3$|9|(\hyperlink{cite.2019ApJ...880....2W}{25}), (\hyperlink{cite.2023ApJ...946L..45T}{39})|$25_{-4}^{+4}$|$46_{-2}^{+3}$|$45_{-2}^{+2}$|$2.3_{-0.2}^{+0.2}$|$40_{ -3 }^{ +3 }$|$59_{-7}^{+9}$
HFLS3|SFG|6.3369|$3.7 \cdot 10^{10}$|$4.9$|28|(\hyperlink{cite.2013Natur.496..329R}{10})|$230_{-8}^{+8}$|$64_{-1}^{+1}$|$39_{-1}^{+1}$|$1.81_{-0.04}^{+0.04}$|$147_{ -3 }^{ +3 }$|$414_{-17}^{+17}$
J1148+5251|QSO|6.4189|nan|$7.1$|8|(\hyperlink{cite.2003A&A...406L..55B}{1}), (\hyperlink{cite.2004MNRAS.351L..29R}{2}), (\hyperlink{cite.2006ApJ...642..694B}{3}), (\hyperlink{cite.2009ApJ...703.1338R}{6}), (\hyperlink{cite.2014MNRAS.445.2848G}{11}), (\hyperlink{cite.2015A&A...574A..14C}{13}), (\hyperlink{cite.2019MNRAS.489.3939C}{20})|$33_{-6}^{+7}$|$64_{-6}^{+7}$|$54_{-5}^{+5}$|$1.7_{-0.2}^{+0.2}$|$37_{ -4 }^{ +5 }$|$241_{-60}^{+76}$
SPT0311-58W|SFG|6.9|nan|$78.5$|7|(\hyperlink{cite.2018Natur.553...51M}{19}), (\hyperlink{cite.2021ApJ...921...97J}{35})|$179_{-35}^{+46}$|$46_{-3}^{+4}$|$40_{-2}^{+2}$|$1.4_{-0.1}^{+0.2}$|$18_{ -2 }^{ +3 }$|$227_{-27}^{+28}$
SPT0311-58E|SFG|6.9|$3.5_{-1.5}^{+1.5} \cdot 10^{10}$|$78.5$|7|(\hyperlink{cite.2018Natur.553...51M}{19}), (\hyperlink{cite.2021ApJ...921...97J}{35})|$37_{-7}^{+9}$|$40_{-3}^{+3}$|$40_{-2}^{+2}$|$2.0_{-0.2}^{+0.2}$|$15_{ -1 }^{ +2 }$|$35_{-4}^{+4}$
A1689-zD1|SFG|7.13|$1.7_{-0.5}^{+0.7} \cdot 10^{9}$|$1.9$|4|(\hyperlink{cite.2015Natur.519..327W}{14}), (\hyperlink{cite.2017MNRAS.466..138K}{16}), (\hyperlink{cite.2020MNRAS.495.1577I}{27}), (\hyperlink{cite.2021MNRAS.508L..58B}{33}), (\hyperlink{cite.2022ApJ...934...64A}{36})|$1.9_{-0.6}^{+1.0}$|$42_{-5}^{+6}$|$39_{-4}^{+4}$|$1.8_{-0.2}^{+0.2}$|$17_{ -3 }^{ +5 }$|$1.9_{-0.4}^{+0.5}$
GNz7q|QSO|7.1899|$2.5_{-1.4}^{+1.4} \cdot 10^{10}$|$0.7$|4|(\hyperlink{cite.2022Natur.604..261F}{37})|$6_{-2}^{+4}$|$64_{-11}^{+12}$|$51_{-10}^{+10}$|$1.8_{-0.2}^{+0.2}$|$56_{ -11 }^{ +18 }$|$35_{-18}^{+30}$
J1342+0928|QSO|7.54|nan|$9.4$|7|(\hyperlink{cite.2019ApJ...881...63N}{23}), (\hyperlink{cite.2020ApJ...904..130V}{32})|$4_{-2}^{+4}$|$56_{-13}^{+23}$|$55_{-11}^{+20}$|$1.9_{-0.2}^{+0.2}$|$12_{ -4 }^{ +5 }$|$26_{-14}^{+52}$
\end{filecontents*}
\csvreader[separator=pipe, late after line=\\, head to column names]{FIR_SED_results.csv}{}{\object & \type & \z & \ifcsvstrcmp{\Mstar}{nan}{\dots}{\Mstar} & \Ad & \nphot & \citations}
            \hline
        \end{tabular}
        \flushleft
        \textbf{Notes.} Listed properties are their type, redshift ($z$), estimates of the stellar mass ($M_*$) and measured deconvolved area of the dust emission ($A_\text{dust}$), number of photometric detections used for the fitting routine ($n_\text{phot}$), and references to the works from which we acquired the FIR photometry.
        References: (\hyperlink{cite.2003A&A...406L..55B}{1}): \citet{2003A&A...406L..55B}, (\hyperlink{cite.2004MNRAS.351L..29R}{2}): \citet{2004MNRAS.351L..29R}, (\hyperlink{cite.2006ApJ...642..694B}{3}): \citet{2006ApJ...642..694B}, (\hyperlink{cite.2006MNRAS.370.1185P}{4}): \citet{2006MNRAS.370.1185P}, (\hyperlink{cite.2007ApJ...671.1531Y}{5}): \citet{2007ApJ...671.1531Y}, (\hyperlink{cite.2009ApJ...703.1338R}{6}): \citet{2009ApJ...703.1338R}, (\hyperlink{cite.2011ApJ...740...63C}{7}): \citet{2011ApJ...740...63C}, (\hyperlink{cite.2011ApJ...740L..15M}{8}): \citet{2011ApJ...740L..15M}, (\hyperlink{cite.2013MNRAS.436.1919C}{9}): \citet{2013MNRAS.436.1919C}, (\hyperlink{cite.2013Natur.496..329R}{10}): \citet{2013Natur.496..329R}, (\hyperlink{cite.2014MNRAS.445.2848G}{11}): \citet{2014MNRAS.445.2848G}, (\hyperlink{cite.2014A&A...569A..98T}{12}): \citet{2014A&A...569A..98T}, (\hyperlink{cite.2015A&A...574A..14C}{13}): \citet{2015A&A...574A..14C}, (\hyperlink{cite.2015Natur.519..327W}{14}): \citet{2015Natur.519..327W}, (\hyperlink{cite.2016MNRAS.457.4406A}{15}): \citet{2016MNRAS.457.4406A}, (\hyperlink{cite.2017MNRAS.466..138K}{16}): \citet{2017MNRAS.466..138K}, (\hyperlink{cite.2017ApJ...845..138S}{17}): \citet{2017ApJ...845..138S}, (\hyperlink{cite.2018A&A...619A..39F}{18}): \citet{2018A&A...619A..39F}, (\hyperlink{cite.2018Natur.553...51M}{19}): \citet{2018Natur.553...51M}, (\hyperlink{cite.2019MNRAS.489.3939C}{20}): \citet{2019MNRAS.489.3939C}, (\hyperlink{cite.2019ApJ...887...55C}{21}): \citet{2019ApJ...887...55C}, (\hyperlink{cite.2019ApJ...877...45M}{22}): \citet{2019ApJ...877...45M}, (\hyperlink{cite.2019ApJ...881...63N}{23}): \citet{2019ApJ...881...63N}, (\hyperlink{cite.2019ApJ...876...99S}{24}): \citet{2019ApJ...876...99S}, (\hyperlink{cite.2019ApJ...880....2W}{25}): \citet{2019ApJ...880....2W}, (\hyperlink{cite.2020ApJ...898...33C}{26}): \citet{2020ApJ...898...33C}, (\hyperlink{cite.2020MNRAS.495.1577I}{27}): \citet{2020MNRAS.495.1577I}, (\hyperlink{cite.2020ApJ...890..171J}{28}): \citet{2020ApJ...890..171J}, (\hyperlink{cite.2020MNRAS.498.4109J}{29}): \citet{2020MNRAS.498.4109J}, (\hyperlink{cite.2020ApJ...889..162L}{30}): \citet{2020ApJ...889..162L}, (\hyperlink{cite.2020ApJ...895...81R}{31}): \citet{2020ApJ...895...81R}, (\hyperlink{cite.2020ApJ...904..130V}{32}): \citet{2020ApJ...904..130V}, (\hyperlink{cite.2021MNRAS.508L..58B}{33}): \citet{2021MNRAS.508L..58B}, (\hyperlink{cite.2021ApJ...923..215C}{34}): \citet{2021ApJ...923..215C}, (\hyperlink{cite.2021ApJ...921...97J}{35}): \citet{2021ApJ...921...97J}, (\hyperlink{cite.2022ApJ...934...64A}{36}): \citet{2022ApJ...934...64A}, (\hyperlink{cite.2022Natur.604..261F}{37}): \citet{2022Natur.604..261F}, (\hyperlink{cite.2022A&A...665A.107T}{38}): \citet{2022A&A...665A.107T}, (\hyperlink{cite.2023ApJ...946L..45T}{39}): \citet{2023ApJ...946L..45T}.
        \label{tab:High-redshift_sources}
    \end{table*}
\endgroup

\section{Methods}
\label{sec:Methods}

We consider spectroscopically confirmed galaxies which are confidently detected in at least four rest-frame FIR photometric bands and have a (deconvolved) dust-continuum size measurement; for the purpose of fitting greybody curves to the dust emission, we consider a rest-frame FIR wavelength range of $10 \, \mathrm{\upmu m} \lesssim \lambda_\text{emit} \lesssim 10^3 \, \mathrm{\upmu m}$. Nearby galaxies used for comparison are discussed in \cref{ssec:Methods:Samples_in_the_local_Universe}, while \cref{ssec:Methods:High-redshift_sample} describes the sources in the early Universe ($4 < z < 8$). We stress that caution has to be taken in interpreting the properties of our sample due to its diversity, as will be discussed in more detail.

\subsection{Samples in the local Universe}
\label{ssec:Methods:Samples_in_the_local_Universe}

The JCMT dust and gas In Nearby Galaxies Legacy Exploration (JINGLE) survey observed $\num{192}$ nearby galaxies with \textit{Herschel} \citep{2018MNRAS.481.3497S, 2019MNRAS.486.4166S, 2019MNRAS.489.4389L}. We make use of photometry at $100 \, \mathrm{\upmu m}$ and $160 \, \mathrm{\upmu m}$ (taken by \textit{Herschel}/PACS) and at $250 \, \mathrm{\upmu m}$, $350 \, \mathrm{\upmu m}$ and $500 \, \mathrm{\upmu m}$ (\textit{Herschel}/SPIRE), excluding the $22 \, \mathrm{\upmu m}$, $60 \, \mathrm{\upmu m}$ and $850 \, \mathrm{\upmu m}$ bands given potential mid-infrared (MIR) excess due to hot dust surrounding active galactic nuclei (AGN) and a non-thermal contribution to the low-frequency emission \citep[some galaxies are not classified as purely star-forming, however for simplicity we apply this label since the sample does not contain bright AGN; see][]{2018MNRAS.481.3497S}. Following \citet{2019MNRAS.489.4389L}, we discard JINGLE-62 given its non-detection at $250 \, \mathrm{\upmu m}$.

A second survey, the \textit{Herschel} (U)LIRG Survey (HERUS), comprises $\num{43}$ nearby (ultra-)luminous infrared galaxies or (U)LIRGs observed by the \textit{Infrared Astronomical Satellite} (\textit{IRAS}) at $100 \, \mathrm{\upmu m}$ and $160 \, \mathrm{\upmu m}$ \citep{2003AJ....126.1607S}, followed up by \textit{Herschel}/SPIRE in the $250 \, \mathrm{\upmu m}$, $350 \, \mathrm{\upmu m}$ and $500 \, \mathrm{\upmu m}$ bands \citep{2018MNRAS.475.2097C}. We exclude 3C273 and IRAS~13451+1232 whose photometric measurements are not fit well by a greybody SED.

Thirdly, we consider a sample of nearby quasars (QSOs) selected from the Palomar-Green (PG) survey \citep{2015ApJS..219...22P}. We use the available far-infrared photometry at $70 \, \mathrm{\upmu m}$, $100 \, \mathrm{\upmu m}$, and $160 \, \mathrm{\upmu m}$ (taken by \textit{Herschel}/PACS) and at $250 \, \mathrm{\upmu m}$, $350 \, \mathrm{\upmu m}$ and $500 \, \mathrm{\upmu m}$ (\textit{Herschel}/SPIRE).

Due to the low spatial resolution of \textit{Herschel}, the size of the dust-continuum emission ($A_\text{dust}$, discussed further in \cref{ssec:Methods:High-redshift_sample,ssec:Methods:Dust_SED_fitting_procedure}) in the samples of local objects is assumed to be equal to the optical size of the galaxy, if available. The properties of all three samples are summarised in \cref{tab:Local_sources}.
\begingroup
    \setlength{\tabcolsep}{10pt} 
    \renewcommand{\arraystretch}{1.25} 
    \begin{table*}
        \centering
        \caption[Results from the \program{mercurius} greybody fitting procedure.]
        {Results from the \program{mercurius} greybody fitting procedure of high-redshift sources discussed in \cref{ssec:Methods:Dust_SED_fitting_procedure}.}
        \begin{tabular}{lllllll}
            Source & $M_\text{dust} \, (10^7 \, \mathrm{M_\odot})$ & $T_\text{dust} \, (\mathrm{K})$ & $T_\text{peak} \, (\mathrm{K})$ & $\beta_\text{IR}$ & $\lambda_0 \, (\mathrm{\upmu m})$ & $L_\text{IR} \, (10^{11} \, \mathrm{L_\odot})$ \\
            \hline
            \csvreader[separator=pipe, late after line=\\, head to column names]{FIR_SED_results.csv}{}{\object & \Md & \Td & \Tp & \betaIR & \lam & \LIR}
            \hline
        \end{tabular}
        \flushleft
        \textbf{Notes.} Columns show the derived dust mass ($M_\text{dust}$), dust temperature ($T_\text{dust}$), peak temperature ($T_\text{peak}$), dust emissivity index ($\beta_\text{IR}$), opacity transition wavelength ($\lambda_0$), and total IR luminosity ($L_\text{IR}$; between $8$ and $\num{1000} \, \mathrm{\upmu m}$). Reported quantities are given as the median (i.e. \nth{50} percentile) of the parameter's marginalised posterior distribution, with a $\pm 1 \sigma$ confidence range reflecting the \nth{16} and \nth{84} percentiles.
        \label{tab:FIR_SED_results}
    \end{table*}
\endgroup
\begin{figure*}
    \centering
    \includegraphics[width=\linewidth]{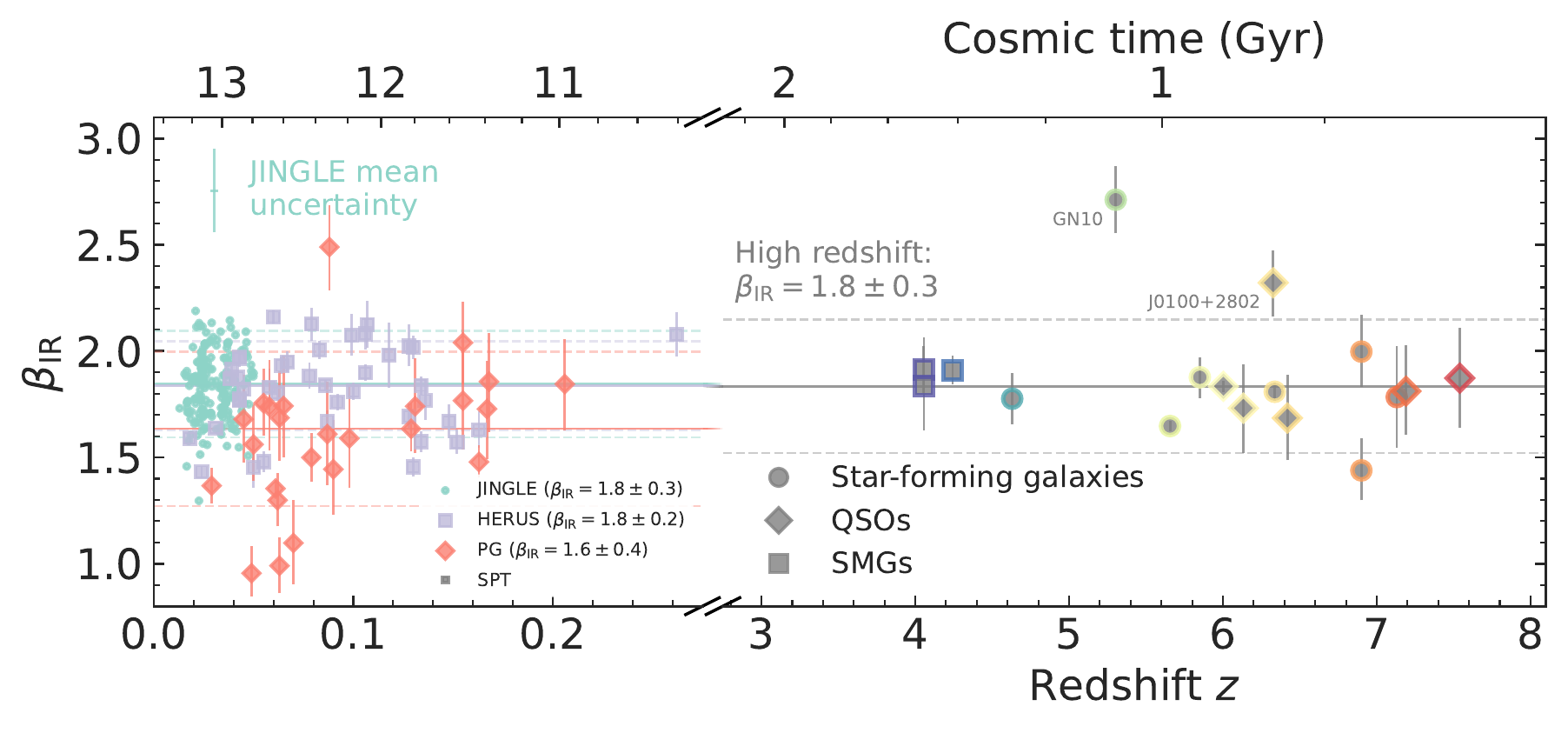}
    \caption[Cosmic evolution of the dust emissivity index $\beta_\text{IR}$.]{Cosmic evolution of the dust emissivity index $\beta_\text{IR}$. The sample of high-redshift galaxies (coloured according to redshift) is described in \cref{ssec:Methods:High-redshift_sample}. At late times, results are shown for JINGLE galaxies (whose mean uncertainty is indicated instead of individual error bars for visualisation purposes), (U)LIRGs from the HERUS survey, and QSOs from the PG sample (\cref{ssec:Methods:Samples_in_the_local_Universe}). 
    Solid (dashed) horizontal lines indicate each sample's mean dust emissivity index (and scatter), consistent across JINGLE, HERUS, and high-redshift samples ($\beta_\text{IR} \simeq 1.8$) while slightly lower for the PG sample ($\beta_\text{IR} \simeq 1.6$).
    }
    \label{fig:Redshift_beta}
\end{figure*}

\subsection{High-redshift sample}
\label{ssec:Methods:High-redshift_sample}

We briefly summarise all 17 high-redshift galaxies and the literature works from which we acquired their compiled FIR SEDs in \cref{tab:High-redshift_sources}. The sources include 6 QSOs, 8 star-forming galaxies (SFG), and 3 sub-millimetre galaxies (SMGs). Size measurements of the dust-continuum emission ($A_\text{dust}$), used to construct a self-consistent opacity model,\footnote{Throughout this work, any discussion of opacity (and optically thin or thick cases) refers to the optical depths in the far-infrared unless specifically stated otherwise.} are taken by combining all reported deconvolved sizes. As will be discussed in \cref{ssec:Methods:Dust_SED_fitting_procedure}, we note these sizes are merely used to inform a first-order approximation of the optical depth as these necessarily correspond to slightly different wavelengths in the rest frame \citep[which, as a result of non-uniform temperature distributions, indeed impacts the measured size; e.g.][]{2022MNRAS.515.1751W, 2022ApJ...934...64A}. We note due to our requirement of having a dust-continuum size measurement, we do not include sources from the South Pole Telescope (SPT) survey \citep{2020ApJ...902...78R} except SPT0346-52 and SPT0311-58 which have been observed separately \citep{2016MNRAS.457.4406A, 2018Natur.553...51M, 2020MNRAS.498.4109J, 2021ApJ...921...97J}. We do however consider the other SPT sources in our discussion of dust temperatures (\cref{ssec:Dust_temperatures}). Similar to the JINGLE and HERUS photometry (\cref{ssec:Methods:Samples_in_the_local_Universe}), we excluded by visual inspection among the high-redshift sample those MIR and radio photometric measurements that clearly have a non-thermal origin and go beyond a simple modified-blackbody SED (see \cref{ap:FIR_SED_fits}, where we present the SEDs). Estimates of the stellar mass ($M_*$), if available, were derived from SED fitting in the works listed under a \citet{2003PASP..115..763C} initial mass function (IMF), except for J2310+1855 and GNz7q where they represent the dynamical mass minus the gas mass, $M_* \simeq M_\text{dyn} - M_\text{gas}$. The number of photometric detections, $n_\text{phot}$, does not count upper limits (or data points that were excluded, as explained in \cref{ssec:Methods:Dust_SED_fitting_procedure}).

\begin{figure*}
	\centering
	\includegraphics[width=\linewidth]{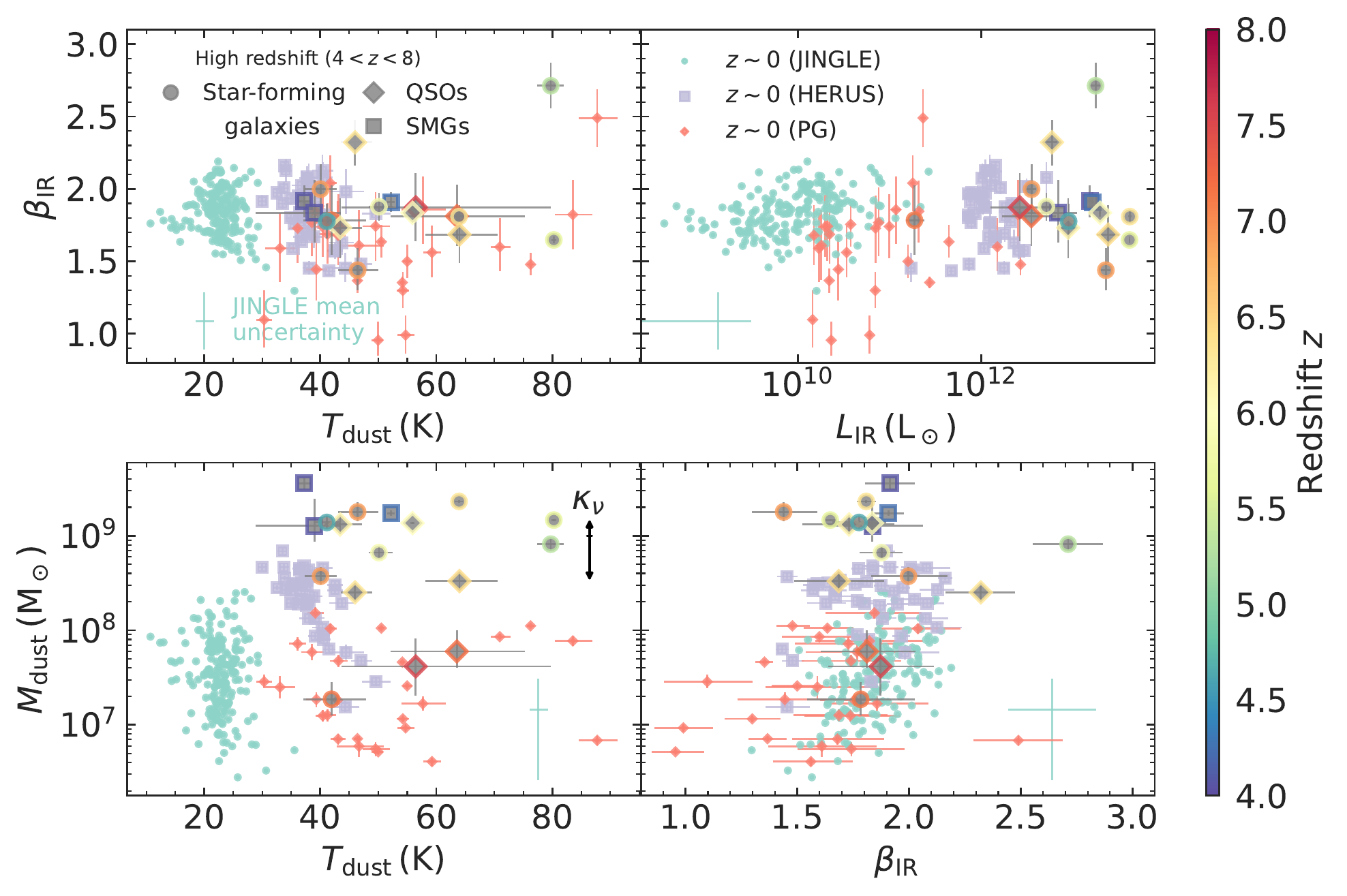}
	\caption[Interdependencies of several dust properties.]{Interdependencies of several dust properties: the dust emissivity index $\beta_\text{IR}$, temperature ($T_\text{dust}$), mass ($M_\text{dust}$), and the total infrared luminosity ($L_\text{IR}$). Data points from the high-redshift sample are coloured according to their redshift. In each panel, instead of individual error bars the mean uncertainty of JINGLE galaxies is indicated for visualisation purposes. In the lower left panel, an arrow shows the systematic uncertainty on dust masses given the range of possible dust absorption cross-sections \citep[$\kappa_\nu$; see][]{2022MNRAS.515.1751W}.
	}
	\label{fig:Galaxy_properties_beta}
\end{figure*}

\subsection{Dust SED fitting procedure}
\label{ssec:Methods:Dust_SED_fitting_procedure}

To derive the dust properties of each galaxy in the three samples discussed above, we fit a greybody SED with a self-consistent opacity model to the photometry with the Bayesian code \program{mercurius}, presented in \citet{2022MNRAS.515.1751W}.\footnote{Available at \url{https://github.com/joriswitstok/mercurius/}.} The derived dust properties of the high-redshift sample are shown in \cref{tab:FIR_SED_results}. We present all SEDs and an example of the obtained posterior distributions in \cref{ap:FIR_SED_fits}.\footnote{Posterior distributions for the rest of our sample are presented as Supplementary material.}

Briefly, we performed two types of fits: either in an entirely optically thin scenario (dashed lines) or with a self-consistent general opacity model, where we link the dust mass surface density, derived using the measured deconvolved area corresponding to the dust emission ($A_\text{dust}$; see \cref{tab:High-redshift_sources}), to the optical depth \citep[see][ for details]{2022MNRAS.515.1751W}. We note that for moderate- and low-resolution measurements where $A_\text{dust}$ (even though it is deconvolved) may effectively represent an upper limit, the general opacity model only marginally differs from an optically thin SED due to the low dust surface mass density. A more accurate treatment of the optical depth would require modelling high-resolution images with detailed, three-dimensional radiative transfer models \citep[e.g.][]{2020MNRAS.495.1577I, 2022MNRAS.516.1612H}, which is beyond the scope of this work. The estimated dust emissivity index is not significantly impacted by the choice of opacity model. The dust temperature, and hence the inferred dust mass, however, are more susceptible to changes in the opacity \citep[see e.g.][]{2020A&A...634L..14C, 2020ApJ...895...81R, 2022A&A...665A...3J}. In some cases, the general opacity model indeed produces notably better results than assuming the SED is entirely optically thin (e.g. GN10 and SPT0346-52; see \cref{ap:FIR_SED_fits}). In the latter case, the a-posteriori (AP) transition wavelength inferred is inconsistent with an optically thin scenario \citep[see also][]{2020MNRAS.498.4109J}. For this reason, we opt for the self-consistent treatment as the fiducial model with the exception of the PG sample, where we fix the transition wavelength to $\lambda_0 = 200 \, \mathrm{\upmu m}$ (having verified an optically thin model only has a minor impact on our findings). Our measurement of the dust mass assumes a fixed dust absorption cross-section $\kappa_\nu$ appropriate for dust ejected by SNe after reverse-shock destruction \citep{2014MNRAS.443.1704H}, with the systematic uncertainty that could lower dust masses by $\ssim 3 \times$ or increase them by $\ssim 1.5 \times$.

\begin{figure*}
	\centering
	\includegraphics[width=\linewidth]{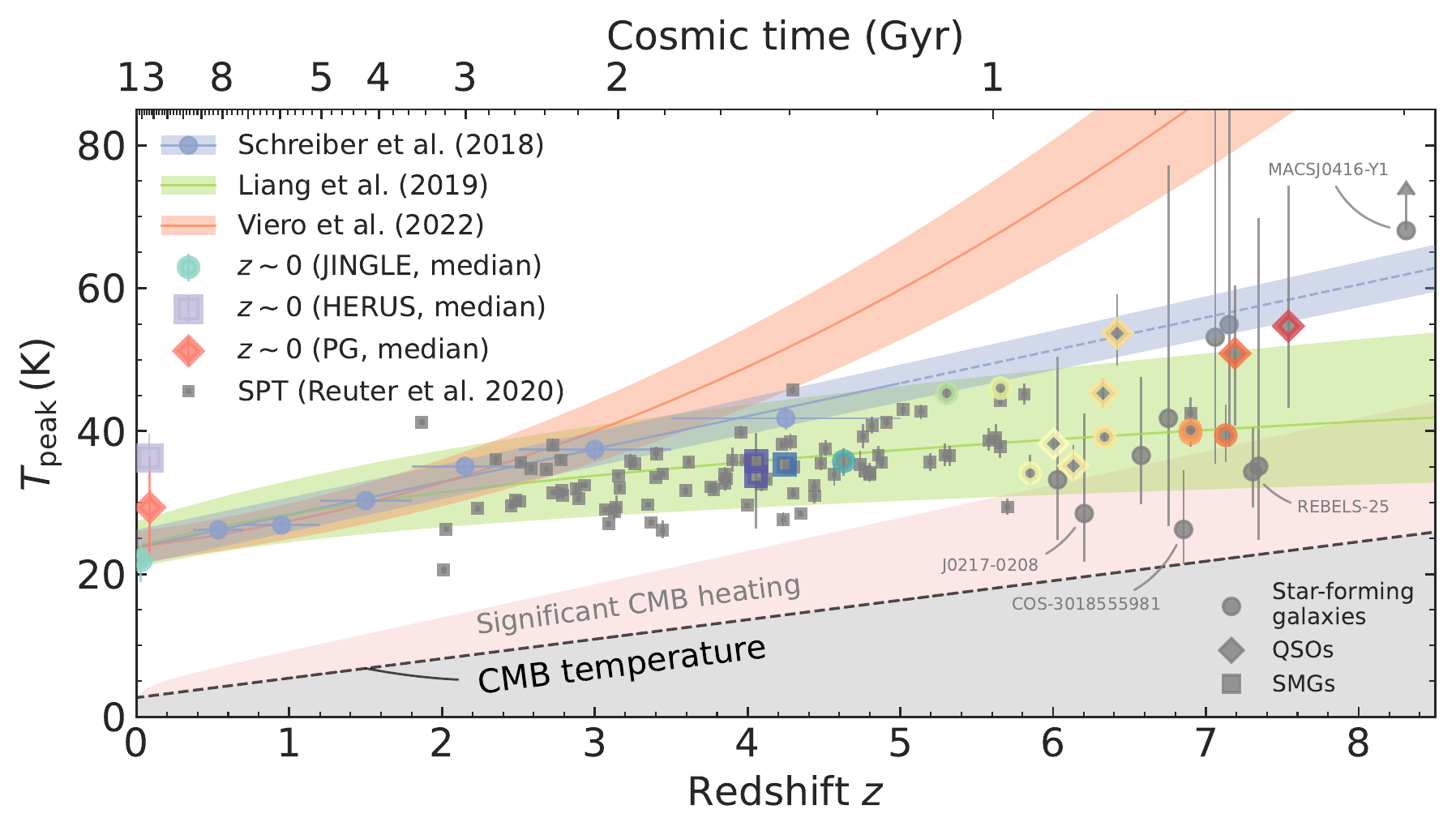}
	\caption[Dust peak temperature as a function of redshift.]{Dust peak temperature, $T_\text{peak}$, as a function of cosmic time. Data points from the high-redshift sample are coloured according to their redshift (symbols are the same as before). Observed trends inferred by \citet{2018A&A...609A..30S} and \citet{2022MNRAS.516L..30V} from stacked spectra are shown, as is the power-law fit to the peak-temperature evolution of simulated galaxies by \citet{2019MNRAS.489.1397L}. Sources from the SPT sample \citep{2020ApJ...902...78R} that have been refitted for consistency (see text for details) are shown as grey squares. We furthermore include several galaxies for which only two to three photometric data points are available to highlight the increased uncertainty (grey points without coloured edge, sources with an extreme dust temperature are annotated; see \cref{ssec:Dust_temperatures} for details). The grey dashed line indicates the CMB temperature, while the light-red area above shows the region where CMB heating is significant \citep[more than $1\%$ difference compared to $z=0$; see][ for details]{2022MNRAS.515.1751W}.
	}
	\label{fig:T_peak}
\end{figure*}

\section{Results and discussion}
\label{sec:Results_and_discussion}

\subsection{Evolution and dependencies of the dust emissivity index at $4 < z < 8$}
\label{ssec:Evolution_of_dust_emissivity}

First, we focus on the dust emissivity index, $\beta_\text{IR}$. We show its redshift evolution in \cref{fig:Redshift_beta}. GN10, with $\beta_\text{IR} \simeq 2.7$, and to a similar extent J0100+2802 with $\beta_\text{IR} \simeq 2.3$, appear to be a clear anomalies compared to the rest of the galaxies analysed here. We note \citet{2020ApJ...895...81R} and \citet{2023ApJ...946L..45T} derive similarly extreme values for the dust emissivity index, which indicates GN10 and J0100+2802 possess exceptional dust properties.

Otherwise, we measure mean dust emissivity index values consistent across the JINGLE, HERUS, and high-redshift samples ($\beta_\text{IR} \simeq 1.8$) while slightly lower for the PG sample ($\beta_\text{IR} \simeq 1.6$). We note that the high-redshift sample will likely have a selection bias towards the brightest IR sources especially because of our requirement of having $4$ confident detections. This may introduce a bias towards low values of $\beta_\text{IR}$ (at fixed dust mass and temperature), however this effect becomes weak at moderately high dust temperatures ($T_\text{dust} \gtrsim 40 \, \mathrm{K}$). To explore further such biases and potential differences between sources at low and high redshift, \cref{fig:Galaxy_properties_beta} shows the interdependencies of several dust properties: the dust emissivity index $\beta_\text{IR}$, temperature ($T_\text{dust}$), mass ($M_\text{dust}$), and the total infrared luminosity ($L_\text{IR}$).

The top left panel illustrates a hint of anti-correlation between the dust emissivity index and temperature among the local samples (mainly in the HERUS sample, $p \approx 0.02$). This well-known degeneracy has been shown to artificially arise from least-squares fitting methods and can be circumvented by hierarchical Bayesian frameworks \citep[e.g.][]{2012ApJ...752...55K, 2019MNRAS.489.4389L}. Indeed, \program{mercurius} is designed to evaluate a likelihood based on the (weighted) sum of squared residuals \citep{2022MNRAS.515.1751W} so that such an artificial spread due to degeneracy may be expected; however, we note it is unclear whether a hierarchical Bayesian approach is appropriate for our high-redshift sample, given its kaleidoscopic nature. Moreover, in this context we are mainly interested in any collective properties of our sample rather than precise individual measurements.

Interestingly, the high-redshift sample best aligns in all panels with the sample of local (U)LIRGs and QSOs, which are offset to higher dust temperatures relative to the JINGLE sample. At the same time, the high-redshift galaxies display more extreme dust masses and IR luminosities than both local galaxy samples. This is certainly (in part) the result of selection bias, since for galaxies in the distant Universe only the bright(er) sources can be discovered and detected across multiple wavelength bands within viable observing times. Still, it is worth noting that already in the first two gigayears of cosmic time these objects are able to form dust masses comparable to, or even larger than, the local galaxies considered here. In the next section, we will discuss the dust temperatures and masses in more detail.

\subsection{Dust temperatures at $4 < z < 8$}
\label{ssec:Dust_temperatures}

The dust temperature has a large impact on derived IR luminosities and SFRs, but it can only be constrained with multi-band FIR observations \citep[e.g.][]{2020MNRAS.498.4192F, 2021MNRAS.508L..58B, 2022MNRAS.515.1751W, 2023arXiv230109659A}. For this reason, both measurements and models of the dust temperature in the early Universe have been scrutinised in recent studies \citep[e.g.][]{2020ApJ...902..112B, 2022MNRAS.510.5560S, 2022MNRAS.513.5621P, 2022MNRAS.511.4999V}. A major hurdle in reconciling theory and observations is the fact that except for an optically thin SED with $\beta_\text{IR} = 2$, the observed peak temperature $T_\text{peak}$ is not trivially related to the dust temperature inferred from the SED, $T_\text{dust}$ \citep[let alone the luminosity- or mass-weighted dust temperature; e.g.][]{2019MNRAS.489.1397L}. In general, disentangling the effects of varying the dust temperature $T_\text{dust}$ and opacity for an optically thick SED requires the SED to be probed at multiple wavelengths \citep{2014PhR...541...45C}.

To explore evolutionary trends in the dust temperature, \cref{fig:T_peak} shows the observed peak temperature as a function of redshift for our sample of sources with well-sampled SEDs, whose temperature can therefore be accurately determined. To aid comparison between samples, we show empirical peak temperatures instead of the more opacity-dependent SED temperature $T_\text{dust}$ \citep[e.g.][]{2014PhR...541...45C, 2020A&A...634L..14C}. We also include several galaxies to highlight the increased uncertainty when only two to three photometric data points are available (all temperatures are inferred consistently with the fitting procedure described in \cref{ssec:Methods:Dust_SED_fitting_procedure}). These are MACSJ0416-Y1 \citep{2019ApJ...874...27T, 2020MNRAS.493.4294B}, B14-65666 \citep{2018MNRAS.481.1631B, 2019PASJ...71...71H, 2021ApJ...923....5S}, J1211-0118, J0217-0208 \citep{2020ApJ...896...93H}, COS-3018555981, UVISTA-Z-001, UVISTA-Z-019 \citep{2022MNRAS.515.1751W}, REBELS-12, REBELS-25, and REBELS-38 \citep{2022MNRAS.515.3126I, 2023arXiv230109659A}. Observed trends inferred by \citet{2018A&A...609A..30S} and \citet{2022MNRAS.516L..30V} from stacked spectra are shown (we note the latter represents the SED temperature $T_\text{dust}$ inferred using an optically thin SED), as is the power-law fit to the peak-temperature evolution of simulated galaxies \citep{2019MNRAS.489.1397L}. In addition, we include sources from the SPT survey (\cref{ssec:Methods:High-redshift_sample}) in this comparison, since the emergent peak temperature can be determined directly from the observed SED and does not strongly depend on the underlying dust opacity \citep[see e.g.][]{2014PhR...541...45C}.\footnote{We note the same does not hold for the SED dust temperature, $T_\text{dust}$: two sources, GN10 and SPT0346-52, are fitted to have an intrinsically high dust temperature (\cref{fig:Galaxy_properties_beta}) while the peak temperature is moderate, owing to their high optical depth.} For consistency, however, we refitted these SEDs with an optically thin SED model, noting there is little difference in the inferred peak temperatures.

Generally, the sources appear to follow the mild temperature evolution found by \citet{2019MNRAS.489.1397L} reasonably well. As is clear from \cref{fig:T_peak}, this is in agreement with trends reported in previous studies of the SPT sources \citep{2020ApJ...902...78R}. Furthermore, a mildly rising temperature it is in line with an evolving main sequence in combination with a non-evolving $L_\text{IR}$-$T_\text{peak}$ relation, as demonstrated at $0 < z < 2$ by \citet{2022ApJ...930..142D}. Moreover, from a theoretical perspective is naturally expected that dust temperatures increase towards high redshift as a result of decreasing gas depletion times \citep{2022MNRAS.513.3122S, 2022MNRAS.517.5930S}. A few sources, however, are clearly offset from this relation derived for star-forming galaxies. In particular, MACSJ0416-Y1 appears to contain exceptionally hot dust \citep[see][]{2020MNRAS.493.4294B}, while J0217-0208, COS-3018555981, and REBELS-25 have notably low temperatures \citep[see also][]{2022MNRAS.515.1751W}. (U)LIRGs from the HERUS sample have increased dust temperatures indicative of a more efficient (i.e. higher UV optical depth) and/or additional heating mechanism \citep[e.g.][]{2022MNRAS.513.3122S, 2022MNRAS.516.1612H}. The same is true for QSOs from the PG sample as well as several high-redshift quasars \citep[in agreement with previous studies; e.g.][]{2021ApJ...921...55M, 2022ApJ...927...21W}.

\begin{figure}
	\centering
	\includegraphics[width=\linewidth]{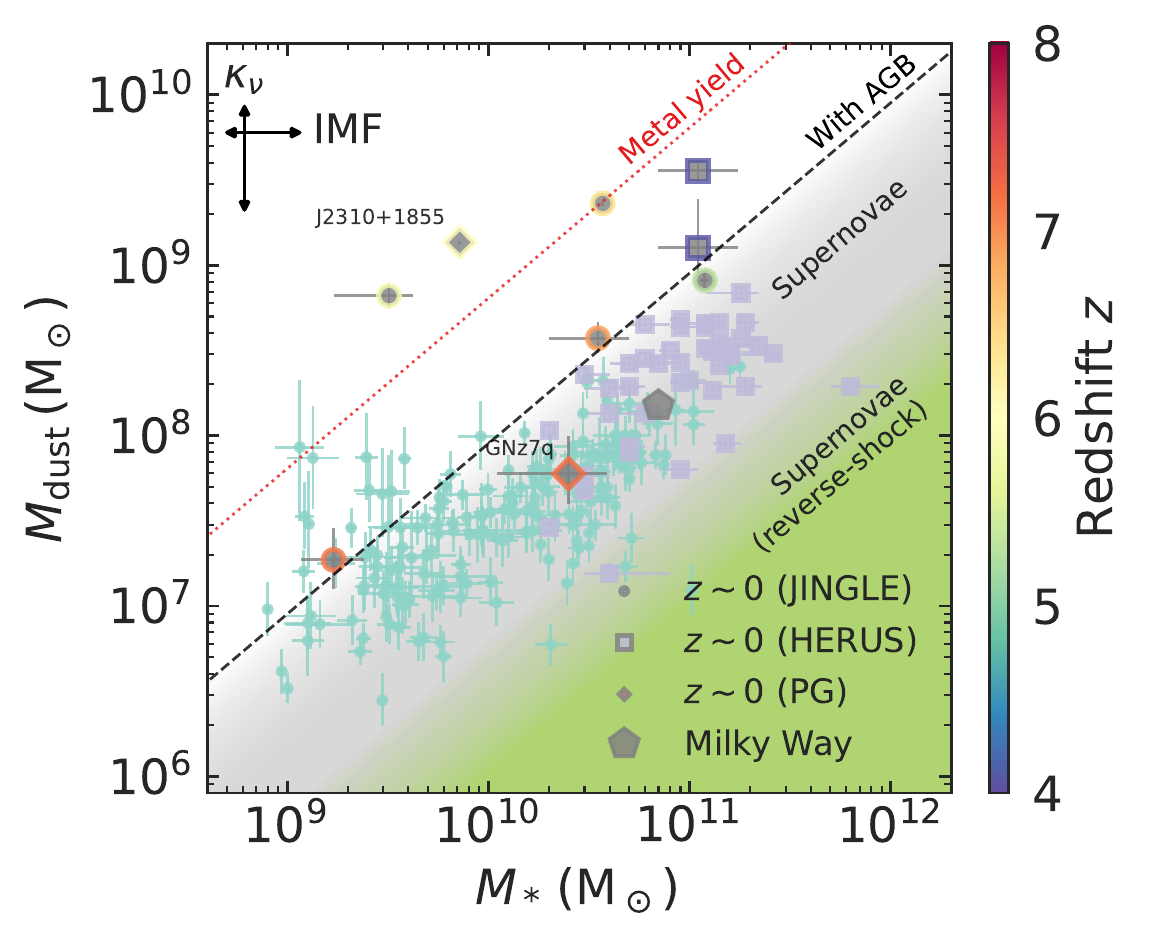}
	\caption[Galaxy dust masses as a function of stellar mass.]{Galaxy dust masses as a function of stellar mass. Data points from the high-redshift sample are coloured according to their redshift (symbols are the same as before). Two QSOs whose stellar mass has been deduced from the dynamical and gas masses (\cref{ssec:Methods:High-redshift_sample}) are annotated. Symbols are the same as in \cref{fig:Redshift_beta,fig:Galaxy_properties_beta}, with the addition of the estimated stellar and dust mass of the Milky Way \citep{2018MNRAS.473.4538G}. Grey shading indicates the region where SNe can be fully responsible for the dust production, green shading shows the same but with a significant reduction due to $\ssim 95\%$ reverse-shock destruction. A maximum yield with additional contribution from AGB stars is indicated by the black dashed line. The edges are blurred on purpose to illustrate the systematic uncertainty in the choice of IMF when calculating the dust yield for a given stellar mass, with the red dotted line showing the highest metal yield, which translates into the maximum dust mass allowed (\cref{ssec:Dust_masses}). As in \cref{fig:Redshift_beta,fig:Galaxy_properties_beta}, an arrow shows the systematic uncertainty on dust masses given the range of possible dust absorption cross-sections (the current choice appropriate for dust ejected by SNe; see \cref{ssec:Dust_masses}), while a second arrow shows the systematic change in observed stellar masses depending on the IMF (see text for details).
	}
	\label{fig:Galaxy_masses}
\end{figure}

\subsection{Dust masses at $4 < z < 8$}
\label{ssec:Dust_masses}

The derived dust masses are shown as a function of the stellar mass (if available; see \cref{ssec:Methods:High-redshift_sample}) in \cref{fig:Galaxy_masses}. In our analysis, we focus on dust production by stellar sources: SN explosions, which are generally thought to be the main stellar source of dust in the early Universe, and AGB stars, which act on longer timescales \citep[e.g.][]{2009MNRAS.397.1661V, 2015MNRAS.451L..70M, 2019MNRAS.490..540L, 2020A&A...637A..32B, 2020A&A...641A.168N}. As outlined in \cref{ssec:Methods:Dust_SED_fitting_procedure}, there is a systematic uncertainty accompanying the choice of the dust absorption cross-section \citep[visualised by an arrow, also shown in \cref{fig:Redshift_beta,fig:Galaxy_properties_beta}; see][]{2022MNRAS.515.1751W}. For consistency, the current choice is appropriate for dust ejected by SNe after reverse-shock destruction \citep{2014MNRAS.443.1704H}.

To relate the stellar and dust masses, we consider an IMF-averaged dust yield $\bar{y}_\text{dust}$, which represents the mean dust mass formed per unit stellar mass and thus translates into a straight line in the dust versus stellar mass plane (i.e. a constant dust-to-stellar mass ratio, $M_\text{dust}/M_*$). Given the uncertainty surrounding theoretical dust yields, in \cref{fig:Galaxy_masses} we indicate a range of dust yields in a scenario where dust is produced by SNe and/or AGB stars under various IMFs following the method described in \citet{2023MNRAS.519.4632D}. Our aim here is to provide a simple prediction of the contribution of stellar sources to the observed dust masses, starting from a given set of theoretically motivated, mass-dependent yields (while varying the stellar IMF). We emphasise this approach differs from the empirical method of \citet{2015A&A...577A..80M} where, inversely, the yield per AGB star or SN event required to explain a given dust mass is inferred \citep[see also][]{2019A&A...624L..13L, 2021MNRAS.508L..58B, 2022MNRAS.515.1751W, 2022ApJ...928...31S, 2022MNRAS.517.5930S}.

The first scenario from \citet{2023MNRAS.519.4632D} assumes a stellar age of $35 \, \mathrm{Myr}$ where only the most massive stars ($M > 8 \, \mathrm{M_\odot}$) contribute to the metal and dust budget through core-collapse SNe. This is shown both for the case where a significant fraction ($\ssim 95\%$) of dust is destroyed by reverse-shock destruction and for the highly optimistic case of no destruction at all \citep[cf.][]{2018MNRAS.473.4538G}. The second has a stellar age of $650 \, \mathrm{Myr}$ (approximately the Hubble time at $z \sim 7$) and includes additional enrichment by lower-mass AGB stars ($M > 2.5 \, \mathrm{M_\odot}$). These two scenarios are chosen to respectively highlight the contribution by SNe alone and the combination of SN and AGB stars. Where available (predominantly in the form of a gas depletion time), estimates of the stellar ages of the high-redshift sources considered in this work are consistently of the order of several tens of millions of years \citep[e.g.][]{2015Natur.519..327W, 2020ApJ...895...81R}, therefore suggesting the dust production by AGB stars is minimal. We note that changing the stellar age only marginally changes the yield: in the first scenario, the most massive stars have smaller ejecta (resulting in a change of $\ssim 10\%$ for an age range of $20$-$35 \, \mathrm{Myr}$), while the AGB contribution in any case is subdominant such that different ages in the second scenario does not significantly change the overall yield. We note that although the timing and grain composition of dust production by SN is uncertain, the adopted yields \citep[$0.16 \, \mathrm{M_\odot}$ and $0.66 \, \mathrm{M_\odot}$;][]{2007MNRAS.378..973B} are within the range of values reported in the literature and certainly are not expected to be significantly higher \citep[e.g.][]{2015A&A...575A..95S, 2016ApJ...817..134L, 2019MNRAS.484.2587M, 2022ApJ...931...85B}.

The gradual diminishing of the shaded area corresponds to the range of yields obtained with various IMFs: a \citet{1955ApJ...121..161S}, \citeauthor{2003PASP..115..763C}, top-heavy ($\alpha = 1.5$), and finally several \citet{1998MNRAS.301..569L} IMFs (with characteristic masses $M_\text{ch}$ of $0.35 \, \mathrm{M_\odot}$, $5 \, \mathrm{M_\odot}$, and $10 \, \mathrm{M_\odot}$; see also \citealt{2009MNRAS.397.1661V}). Yields range from $\bar{y}_\text{SN} = 0.00159$ ($\bar{y}_\mathrm{SN+AGB} = 0.00183$) for a \citeauthor{1955ApJ...121..161S} IMF up to $\bar{y}_\text{SN} = 0.00884$ ($\bar{y}_\mathrm{SN+AGB} = 0.00899$) for a \citeauthor{1998MNRAS.301..569L} IMF ($M_\text{ch} = 10 \, \mathrm{M_\odot}$), which reduces to respectively $\bar{y}_\text{SN} = 1.1 \cdot 10^{{-4}}$ ($\bar{y}_\mathrm{SN+AGB} = 3.46 \cdot 10^{{-4}}$) and $\bar{y}_\text{SN} = 6.22 \cdot 10^{{-4}}$ ($\bar{y}_\mathrm{SN+AGB} = 9.14 \cdot 10^{{-4}}$, but in this case for $M_\text{ch} = 5 \, \mathrm{M_\odot}$) in the presence of a reverse shock causing $\ssim 95\%$ destruction. These yields broadly match more detailed models, which also include the effects of dust reprocessing in the ISM (destruction by SN shocks, astration, ejection in outflows, grain growth). Such analyses predict dust-to-stellar mass ratios depending on stellar mass and redshift, performed in semi-analytical setting \citep[e.g.][]{2015MNRAS.451L..70M, 2017MNRAS.471.3152P, 2019MNRAS.489.4072V, 2022MNRAS.512..989D} and cosmological hydrodynamical simulations \citep[e.g.][]{2017MNRAS.468.1505M, 2018MNRAS.478.4905A, 2020MNRAS.491.3844A, 2020MNRAS.494.1071G, 2023MNRAS.519.4632D}: for instance, \citet{2023MNRAS.519.4632D} predict $M_\text{dust}/M_* \approx 3 \cdot 10^{-4}$ for $M_* \sim 10^9 \, \mathrm{M_\odot}$ to $M_\text{dust}/M_* \approx 10^{-2}$ for $M_* \sim 10^{11} \, \mathrm{M_\odot}$.

From \cref{fig:Galaxy_masses}, it appears local galaxies (mostly) exhibit dust-to-stellar mass ratios consistent with the adopted dust yields, with a possible exception at low stellar masses ($M_* \lesssim 4 \cdot 10^9 \, \mathrm{M_\odot}$) where a small fraction of star-forming galaxies from the JINGLE sample nominally exceed the maximum yield of SN and AGB stars combined. It appears little room is left for a (significant) fraction of SN dust to be subsequently destroyed by the various processes mentioned above, or otherwise that any loss is made up by grain growth \citep[see e.g.][]{2011MNRAS.416.1916V, 2014MNRAS.444.2442V, 2020MNRAS.494.1071G, 2022MNRAS.512..989D}.

The high-redshift galaxies for which a measurement of the stellar mass is available paint a more drastic picture still: a significant fraction are located either in a region where $\ssim 95\%$ reverse-shock destruction is strongly ruled out (i.e. approximately on the black dashed line; 4 out of 9 sources), or indeed where even the optimistic SN and AGB dust yields combined appear inadequate in accounting for the observed dust mass (i.e. significantly above the black dashed line; an additional 4 out of 9 sources). We note that the latter four sources (MORA-4 a.k.a. MAMBO-9, J2310+1855, HFLS3, and GN20) are some of the FIR-brightest and highly starburst (in the case of J2310+1855, quasar-host) galaxies known, likely or demonstrably located within highly overdense regions of the early Universe \citep{2013Natur.496..329R, 2014A&A...569A..98T, 2019ApJ...887...55C, 2022A&A...665A.107T}. In this case, other dust production mechanisms, such as direct grain growth in the ISM, may need to be evoked \citep[see also e.g.][]{2014MNRAS.444.2442V, 2016MNRAS.457.1842S, 2017MNRAS.471.3152P, 2020MNRAS.494.1071G, 2023MNRAS.519.4632D}. However, we note direct grain growth may not adequately resolve the tension as this process has been shown to become inefficient in dense molecular clouds due to the formation of icy mantles \citep{2016MNRAS.463L.112F, 2018MNRAS.476.1371C}. Moreover, two out of the offending data points are incompatible even with the maximum dust mass that is physically allowed -- assuming all metals are locked up in dust grains -- even for the maximum metal yield out of all scenarios described above ($y_Z = 0.0641$ on a longer timescale for a \citeauthor{1998MNRAS.301..569L} IMF with $M_\text{ch} = 10 \, \mathrm{M_\odot}$). We note that although metal and dust yields at a fixed stellar mass are higher for top-heavy IMFs, the observed stellar masses currently inferred with a \citeauthor{2003PASP..115..763C} IMF will also be reduced (as indicated by the arrow in \cref{fig:Galaxy_masses}), again worsening the discrepancy. This suggests that, independent of the IMF assumed, stellar mass estimates may be systematically low, in particular for the high-redshift sources \citep[in agreement with recent findings that suggest the stellar masses can vary up to $\ssim 1 \, \mathrm{dex}$ under a different star formation history; see][]{2022MNRAS.516..975T}. Improved stellar-mass estimates with \textit{JWST} will significantly reduce their uncertainty, though we note high dust yields are not only seen for galaxies whose stellar mass has been measured with SED fitting, but also via the dynamical mass \citep[i.e. for J2310+1855; however, dynamical-mass estimates equally carry significant uncertainty as they require many assumptions to be made in the dynamical modelling and gas mass inference; cf.][]{2018A&A...619A..39F, 2022A&A...665A.107T}.

The inconsistency may therefore also be alleviated with a lower dust absorption cross-section, which directly increases the inferred dust masses. Due to spatial offsets between the light from (unobscured) stars, ionised gas, and dust \citep[e.g.][]{2017A&A...605A..42C, 2018MNRAS.477..552B, 2018MNRAS.481.1631B, 2022MNRAS.510.5088B, 2019MNRAS.488.1779C, 2022MNRAS.515.3126I}, integrated measures such as the stellar mass and dust mass could in reality trace different components. A more heterogeneous dust distribution, for example in a scenario with multiple star-forming regions enshrouded by dust to varying degrees, might bias spatially resolved observations to trace optically thick dust efficiently heated by stars, while colder, possibly more extended dust reservoirs remain undetected \citep[e.g.][]{2017MNRAS.471.5018F, 2022MNRAS.515.1751W, 2022ApJ...934...64A}. High-resolution, multi-wavelength follow-up observations (with e.g. ALMA and \textit{JWST}) are required to confirm this.

\section{Summary and conclusions}
\label{sec:Summary}

We applied a flexible fitting routine to the dust continuum emission of a sample of 17 galaxies in the early Universe ($4 < z < 8$) with well-sampled FIR SEDs, consisting of 6 quasars (QSOs), 8 star-forming galaxies (SFGs), and 3 sub-millimetre galaxies (SMGs). We put the results into context by comparing to two samples of nearby galaxies. We constrain the dust mass, temperature, emissivity index, and opacity of each source within a Bayesian inference framework to characterise collective trends of the dust properties. We summarise our findings as follows:

\begin{itemize}
    \item We find no significant evolution in the dust emissivity index across cosmic time, measuring a consistent value of $\beta_\text{IR} \simeq 1.8$ for the local samples of star-forming galaxies and (U)LIRGs as well as in the high-redshift sample, while it is slightly lower for the sample of local QSOs ($\beta_\text{IR} \simeq 1.6$). This suggests the effective properties of dust do not change dramatically, though we note our selection criteria might introduce a bias towards the IR-brightest sources. GN10 and J0100+2802, with $\beta_\text{IR} \simeq 2.7$ and $\beta_\text{IR} \simeq 2.3$ respectively, are two notable exceptions among the galaxies considered here.
    
    \item In other dust properties, we find the high-redshift galaxies to be similar to, or in dust temperature and IR luminosity even more extreme than, (U)LIRGs in the HERUS sample. The dust temperature in our high-redshift sample points towards a mild evolution with redshift, though several of the most distant quasars, like local quasars and (U)LIRGs, exhibit elevated temperatures pointing towards an additional and/or more efficient heating mechanism.
    
    \item For a subsample of galaxies, we compare stellar-mass estimates with our inferred dust masses, whose degeneracy mainly with dust temperature is reduced for well-constrained SEDs. About half of the cases fall within a simple, IMF-averaged dust yield of stellar sources alone (i.e. SNe and AGB stars), provided no destruction in the ISM or ejection from the galaxy takes place. The other half, however, exceeds a highly optimistic yield by up to an order of magnitude, suggesting stellar masses are underestimated or metal yields are seemingly violated, both implying the mass-to-light ratio of dust may be systematically overestimated.
\end{itemize}

\section*{Data availability}

No new data were generated or analysed in support of this research.

\section*{Acknowledgements}

We are grateful to Stephen Eales and Laura Sommovigo for fruitful discussions. We further thank the anonymous referee for their helpful suggestions. JW, GCJ, RM, and RSm acknowledge support from the ERC Advanced Grant 695671, ``QUENCH''. JW furthermore gratefully acknowledges support from the Fondation MERAC. GCJ acknowledges funding from ERC Advanced Grant 789056 ``FirstGalaxies’’. RM also acknowledges funding from a research professorship from the Royal Society. RSm acknowledges a STFC Ernest Rutherford Fellowship (ST/S004831/1). RSc acknowledges support from the Amaldi Research Center funded by the MIUR program `Dipartimento di Eccellenza' (CUP:B81I18001170001). This work has also used the following packages in \program{python}: the \program{SciPy} library \citep{Jones2001}, its packages \program{NumPy} \citep{2011CSE....13b..22V} and \program{Matplotlib} \citep{Hunter2007}, the \program{Astropy} package \citep{2013A&A...558A..33A, 2018AJ....156..123A}, and the \program{pymultinest} package \citep{2009MNRAS.398.1601F, 2014A&A...564A.125B}.

\bibliographystyle{mnras}
\bibliography{Dust_SED_fitting}

\appendix

\section{FIR SED fits}
\label{ap:FIR_SED_fits}

\begin{figure*}
	\centering
	\includegraphics[width=\linewidth]{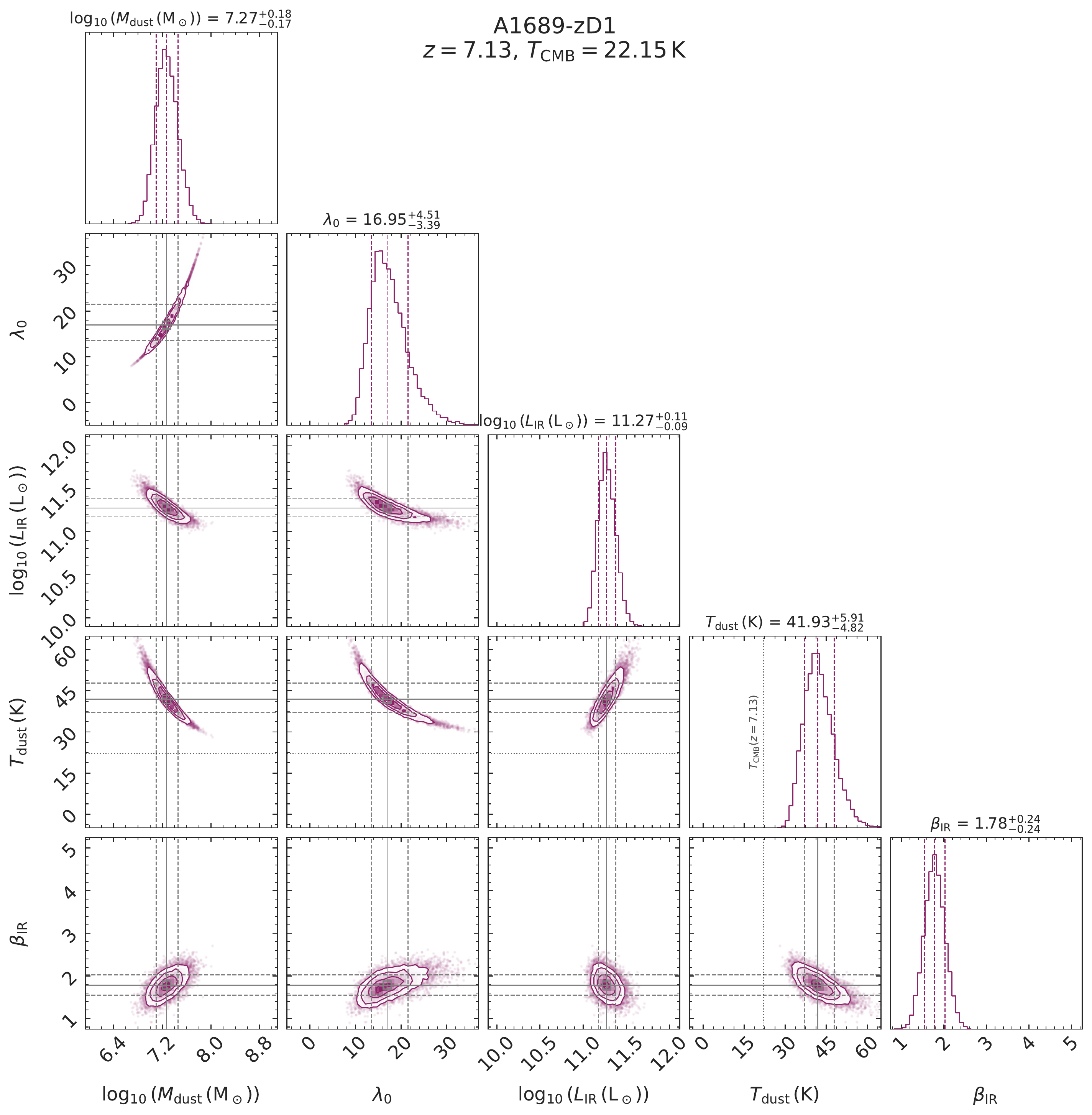}
	\caption[\program{mercurius} posterior distributions for A1689-zD1.]{Posterior distributions for A1689-zD1 at $z = 7.13$ obtained in the FIR SED fitting routine with \program{mercurius} under a self-consistent general opacity model. Parameters shown are the dust mass ($M_\text{dust}$), opacity transition wavelength ($\lambda_0$), total IR luminosity ($L_\text{IR}$; note this is included purely for visualisation, not being an independent parameter in the fitting routine), dust temperature ($T_\text{dust}$), and dust emissivity index ($\beta_\text{IR}$). Solid grey lines indicate the median (i.e. \nth{50} percentile) of the parameter's marginalised posterior distribution, while dashed lines show the \nth{16} and \nth{84} percentiles. In panels with the dust temperature, a dotted line indicates $T_\text{CMB} (z=7.13)$, the CMB temperature at the redshift of A1689-zD1.
	}
	\label{fig:A1689-zD1_SED_posterior}
\end{figure*}

As an example, the posterior distributions for fitting the FIR SED of A1689-zD1 obtained with \program{mercurius} under a self-consistent general opacity model are shown in \cref{fig:A1689-zD1_SED_posterior}. Similar figures for other sources in our sample (\cref{tab:High-redshift_sources}) are included in the Supplementary material.

In \cref{fig:FIR_SED_fits0}, we show the ``best-fit'' SED curve for the maximum likelihood in the $(M_\text{dust}, T_\text{dust})$ plane, while shaded regions indicate the uncertainty as the deviation of the \nth{16} and \nth{84} percentiles from the median (\nth{50} percentile) at each wavelength of all curves produced according to the posterior distribution. We show two types of fits: either in an entirely optically thin scenario (dashed lines) or with a self-consistent general opacity model, where we link the dust mass surface density, derived using the measured deconvolved area corresponding to the dust emission ($A_\text{dust}$; see \cref{tab:High-redshift_sources}), to the optical depth \citep[see][ for details]{2022MNRAS.515.1751W}.

\begin{figure*}
	\centering
	\includegraphics[width=\linewidth]{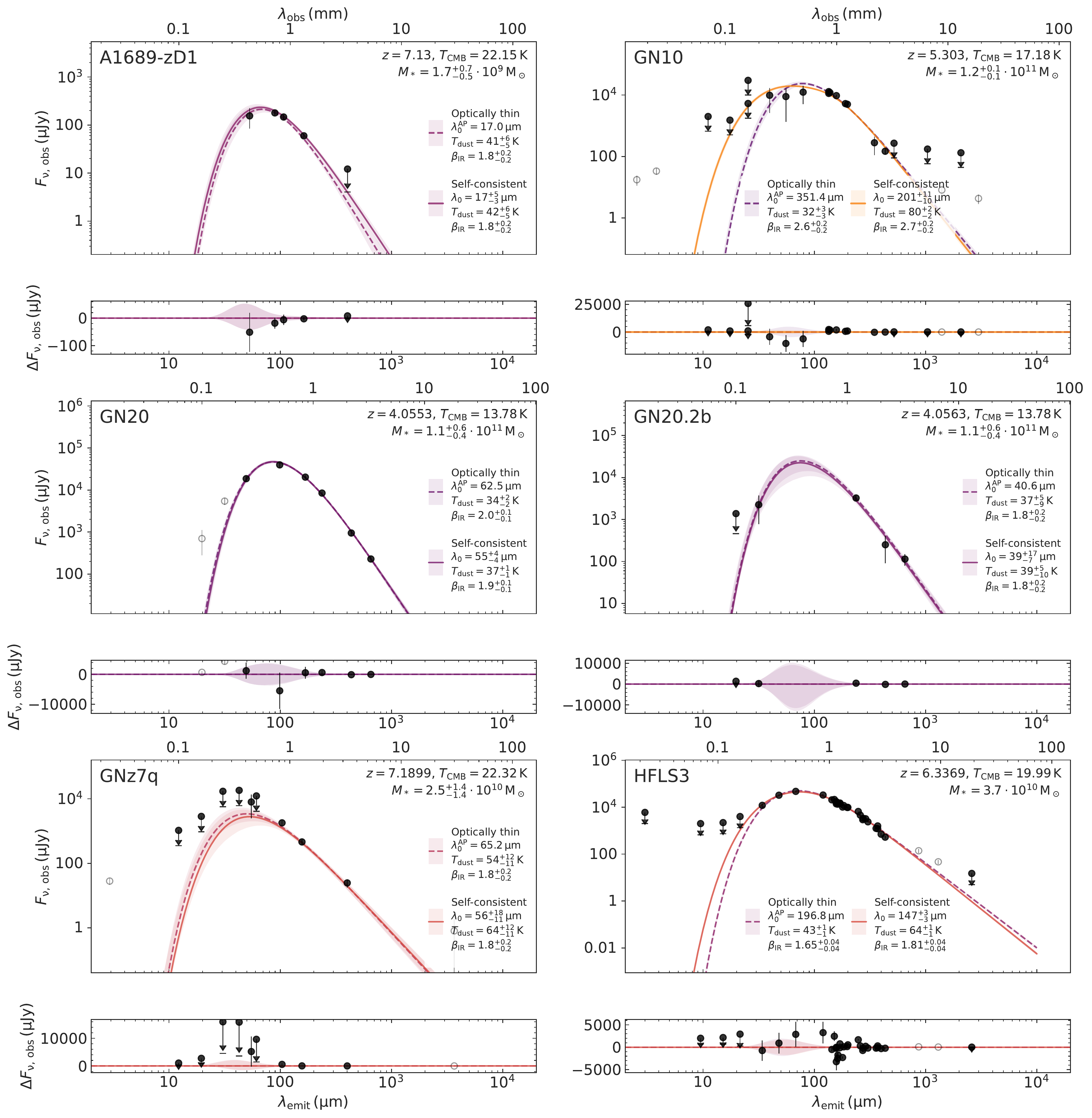}
	\caption[\program{mercurius} fits to the FIR SEDs of the high-redshift sources considered in this work.]{\program{mercurius} fits to the FIR SEDs of the high-redshift sources considered in this work (\cref{tab:High-redshift_sources}). Photometric measurements and upper limits ($3\sigma$, with a vertical dash showing the $1\sigma$ level) are shown in black (open grey points are excluded from the fitting procedure; \cref{ssec:Methods:Dust_SED_fitting_procedure}). Two types of fits are shown, either in an entirely optically thin scenario (dashed lines) or with a self-consistent optical depth model \citep[solid lines; cf.][]{2022MNRAS.515.1751W}. All dust properties derived using the latter model are listed in \cref{tab:FIR_SED_results}. Shaded regions indicate the uncertainty (see text for details). The label lists the transition wavelength ($\lambda_0$), dust temperature ($T_\text{dust}$), and dust emissivity index ($\beta_\text{IR}$). In the optically thin case, the transition wavelength is inferred a posteriori ($\lambda_0^\text{AP}$). A second, smaller panel below each fit shows the residuals ($\Delta F_{\nu, \, \text{obs}}$) on a linear scale.
	}
	\label{fig:FIR_SED_fits0}
\end{figure*}

\begin{figure*}
	\centering
	\includegraphics[width=\linewidth]{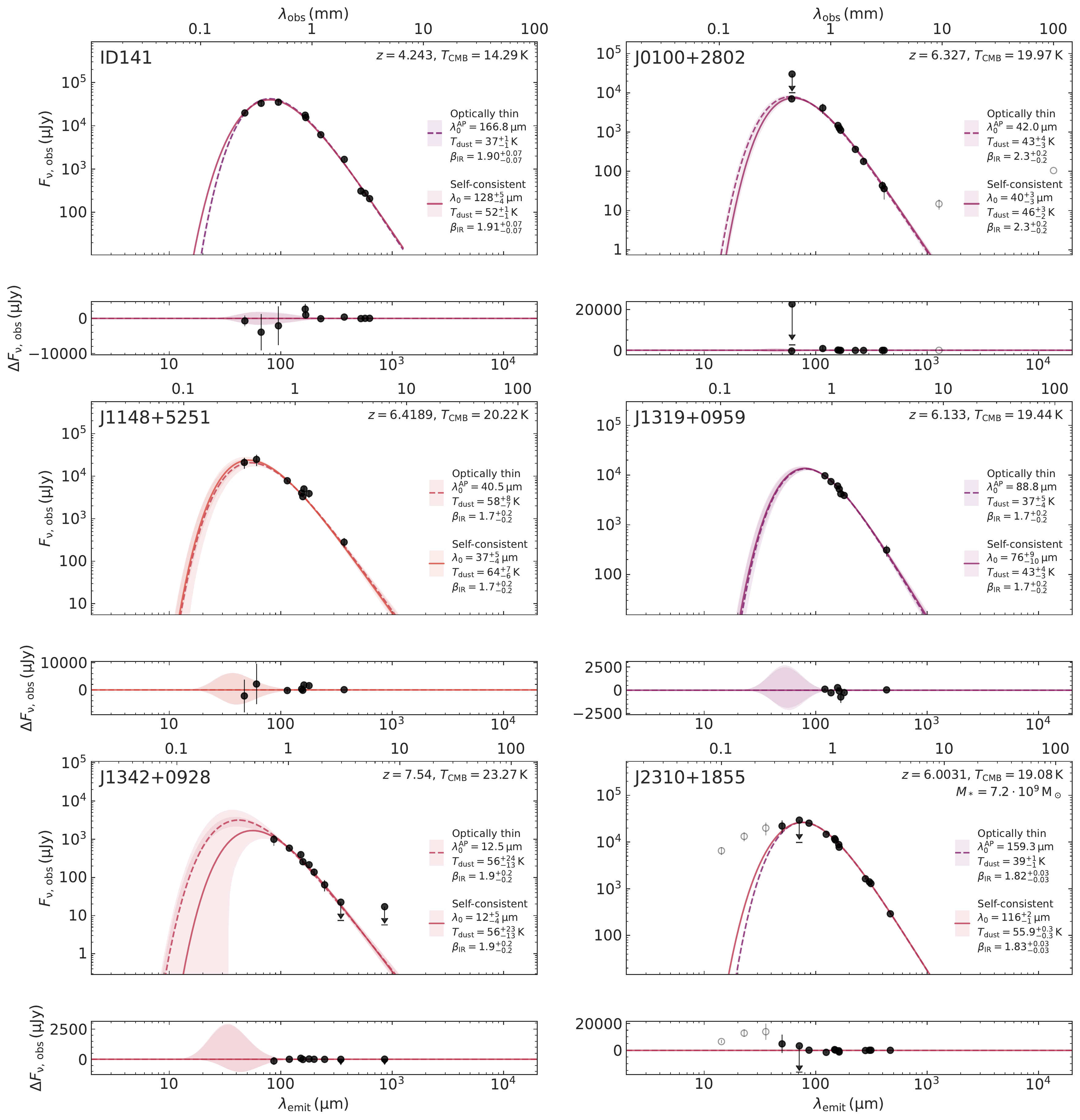}
	\textbf{\cref{fig:FIR_SED_fits0}} (continued).
\end{figure*}

\begin{figure*}
	\centering
	\includegraphics[width=\linewidth]{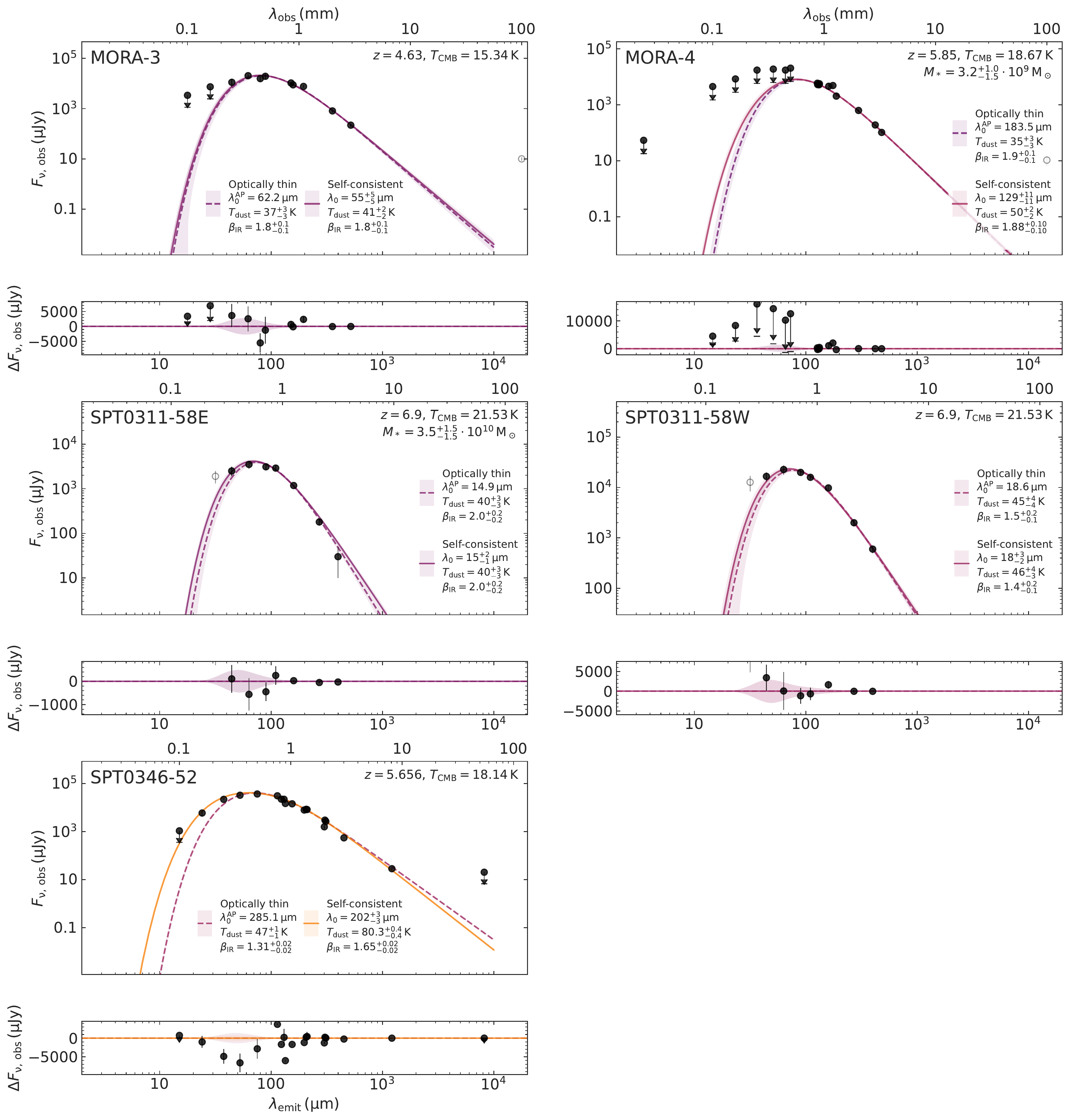}
	\textbf{\cref{fig:FIR_SED_fits0}} (continued).
\end{figure*}


\bsp	
\label{lastpage}
\end{document}